\documentclass{article}

\usepackage[english]{babel}

\usepackage[letterpaper,top=2cm,bottom=2cm,left=3cm,right=3cm,marginparwidth=1.75cm]{geometry}

\usepackage{amsmath}
\usepackage{graphicx}
\usepackage[colorlinks=true, allcolors=blue]{hyperref}

\title{Lateral Ventricle Shape Modeling using Peripheral Area Projection for Longitudinal Analysis}
\author{Wonjung Park, Suhyun Ahn, Jinah Park\\
\normalsize{Korea Advanced Institute of Science and Technology, KAIST}\\\normalsize{fabiola@kaist.ac.kr}}
\date{}
\begin{document}
\maketitle

\begin{abstract}
The deformation of the lateral ventricle (LV) shape is widely studied to identify specific morphometric changes associated with diseases.
Since LV enlargement is considered a relative change due to brain atrophy, local longitudinal LV deformation can indicate deformation in adjacent brain areas.
However, conventional methods for LV shape analysis focus on modeling the solely segmented LV mask.
In this work, we propose a novel deep learning-based approach using peripheral area projection, which is the first attempt to analyze LV considering its surrounding areas.
Our approach matches the baseline LV mesh by deforming the shape of follow-up LVs, while optimizing the corresponding points of the same adjacent brain area between the baseline and follow-up LVs. Furthermore, we quantitatively evaluated the deformation of the left LV in normal (n=10) and demented subjects (n=10), and we found that each surrounding area (thalamus, caudate, hippocampus, amygdala, and right LV) projected onto the surface of LV shows noticeable differences between normal and demented subjects.
\end{abstract}
\section{Introduction}
Enlargement of the lateral ventricles (LV) is observed alongside the loss of brain cells during normal aging and is more pronounced in diseases such as dementia.
Thus, LV shape deformation is widely studied to find specific changes associated with brain atrophy.
The primary method for shape analysis compares baseline and follow-up LV obtained from longitudinal brain images.
Longitudinal changes are derived by deforming the original LV to match the target shape and comparing the original and deformed one.
For example, deformable mesh-based methods~(\cite{spharm}) deform the source LV mesh to the target shape, while diffeomorphic mapping-based methods~(\cite{qiu2008multi, lddmm}) warp the source brain image to align with the target image.

Despite the successful alignment of the LV shapes, previous approaches cannot ensure that the corresponding points on the baseline and follow-up LV surfaces accurately represent local deformations. To be more specific, these methods do not guarantee that points along brain parts such as the amygdala and hippocampus in the baseline LV are moved to the same part in the follow-up LV.

In our work, taking into account that LV shape deformation is accompanied by surrounding brain atrophy, we utilize peripheral area information to find corresponding points more accurately. 
Our deep learning-based method induces the vertices of the source mesh to move to the corresponding points located in the same surrounding brain area.

To demonstrate the quantitative local deformation of the LV while considering peripheral areas, we applied our approach to longitudinal brain MRIs of normal (n=10) and demented (n=10) subjects.
By projecting peripheral brain structures onto the LV, we analyzed local deformations relative to adjacent brain regions.
Noticeable differences in local surface deformations adjacent to the thalamus, caudate, hippocampus, amygdala, and right LV were observed.

\section{Method}
\begin{figure}[t]\centering
\includegraphics[width=0.75\textwidth]{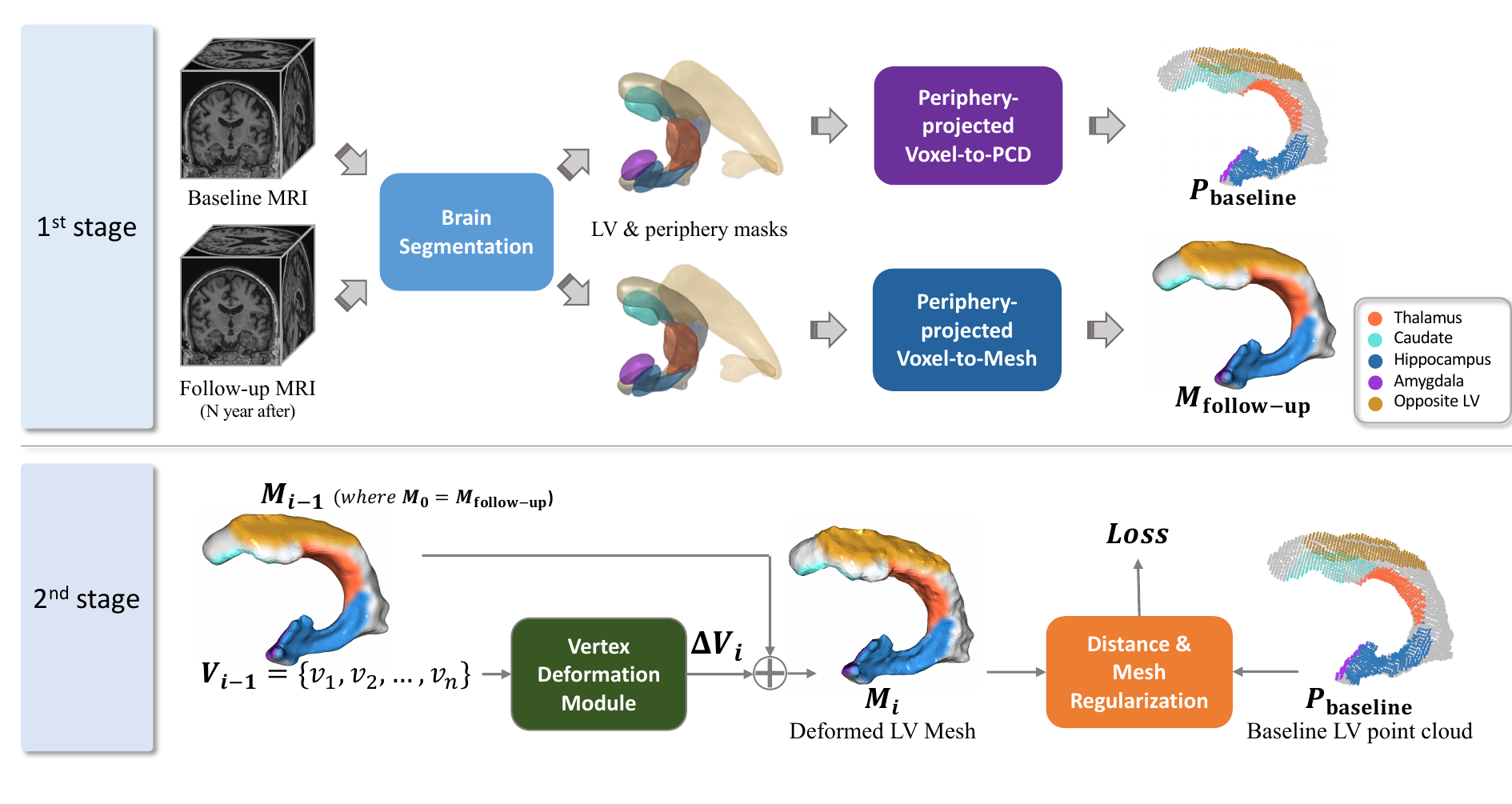}
\caption{\textbf{Overall architecture to estimate longitudinal shape deformation of lateral ventricle.}} \label{fig1}
\end{figure}

As illustrated in Fig.~\ref{fig1}, our method for longitudinal analysis of LVs consists of two stages. In the first stage, the LV and its peripheral areas are segmented from registered longitudinal brain MRIs. Using the LV mask, we constitute point cloud-based baseline and mesh-based follow-up LVs. Then, using surrounding area masks, we classify the points of the baseline point cloud and the vertices of the follow-up mesh into the nearest peripheral areas.

In the second stage, the follow-up mesh is iteratively deformed to match the baseline shape using a vertex deformation module based on Pointnet~(\cite{pointnet}) architecture with the input of the vertex position. The objective function to minimize is the sum of the distance loss ${L}_{dist}$ between the deformed mesh and the baseline point cloud, and the regularization loss ${L}_{reg}$ for the deformed mesh.
The distance loss is:
\begin{equation}
{L}_{dist} = {\lambda}_{cf}{L}_{cf}+{\lambda}_{pm}{L}_{pm}+\sum_{i=1}^{m}{\lambda}_{pm_i}{L}_{pm_i}.
\end{equation}

${L}_{cf}$ is chamfer distance between the selected points on the deformed mesh ${M}_{i}$ and the point cloud ${P}_{baseline}$. ${L}_{pm}$ is bidirectional distance between ${P}_{baseline}$ and the faces of ${M}_{i}$. ${L}_{pm_i}$ represents the bidirectional distance between the points and the part of the mesh faces classified into the $i$-th area of $m$ surrounding area classes. This loss encourages the vertices of the mesh to deform towards the positions of the same surrounding area.
The regularization loss is:
\begin{equation}
{L}_{reg} = {\lambda}_{vert}{L}_{vert}+{\lambda}_{edge}{L}_{edge}+{\lambda}_{normal}{L}_{normal}+{\lambda}_{lap}{L}_{lap}.
\end{equation}
${L}_{vert}$ represents the root mean square distance of the vertices moved, which induces the smallest movements of the vertices as the deformable mesh approaches the target shape. ${L}_{edge}$ is edge length regularization to prevent skewed mesh. Normal consistency ${L}_{normal}$, and Laplacian smoothness ${L}_{lap}$ are used to derive a smooth deformed mesh.



\section{Results}

\begin{figure}[t]\centering
\includegraphics[width=0.8\textwidth]{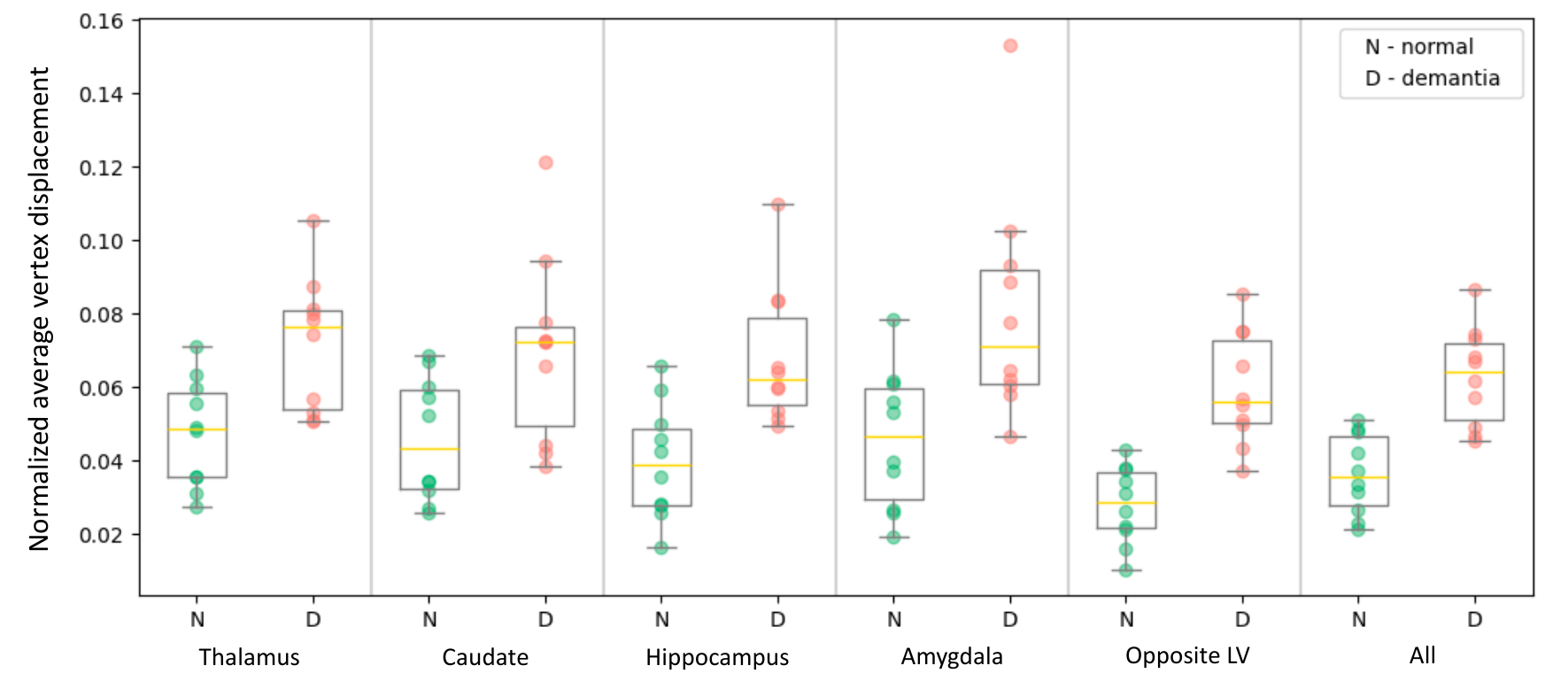}
\caption{Comparison of longitudinal averaged vertex displacement in peripheral brain areas between two groups: normal (n=10) and demented (n=10) subjects. The displacement is normalized by the product of the time interval between two longitudinal MRIs (in year) and one-third power of the individual estimated total intracranial volume.} \label{fig2}
\end{figure}

We analyzed LV shape deformation using our approach in the OASIS longitudinal dataset~(\cite{OASIS}) of normal (n=10) and demented (n=10) subjects whose baseline ages ranged from 78 to 88 years.
To derive the deformation, the distance between the baseline and follow-up shapes is calculated by comparing the original vertex ${V}_{0}$ with the deformed vertex ${V}_{k}$ after k optimization iterations.
As depicted in Fig.~\ref{fig2}, we analyzed changes in regions adjacent to thalamus, caudate, hippocampus, amygdala, and the opposite LV, as well as in the entire LV.
As a result, demented subjects exhibited larger deformations than normal in all cases.
In addition, we performed a comparative analysis between normal male (n=19, age range=80.1$\pm$5.5) and female (n=19, age range=80.6$\pm$5.0) subjects sourced from the OASIS dataset. Across all examined regions, no statistically significant differences in deformation were observed between sexes (p-value>0.26).

\section{Conclusion}
We suggest a deep learning-based LV shape modeling for longitudinal analysis using peripheral area projection that enables the interpretation of local deformations along surrounding areas.
We applied our method to the longitudinal dataset and verified its stable shape recovery while maintaining the locality of the surrounding structures. Furthermore, we demonstrated noticeable differences in deformation patterns between normal and demented subjects.

\section{Acknowledgements}
This work was supported by Institute for Information \& communications Technology Promotion(IITP) grant funded by the Korea government(MSIT) (No.00223446, Development of object-oriented synthetic data generation and evaluation methods) and Korea Center for Gendered Innovations for Science and Technology Research (GISTeR), through the Center for Women in Science, Engineering and Technology (WISET) funded by the Ministry of Science and ICT (WISET202403GI01)

\bibliographystyle{alpha}
\bibliography{sample}

\end{document}